# Quantum chemical studies on beryllium hydride oligomers

**Ch. Bheema Lingam,**[a] **K. Ramesh Babu,**[b] **Surya P. Tewari**[a,b]**, G. Vaitheeswaran**,[b,*]

[a]*School of Physics, University of Hyderabad, Hyderabad-500046, India*

[b]*Advanced Centre of Research in High Energy Materials (ACRHEM), University of Hyderabad, Hyderabad-500046, India.*

## Abstract

The present study explores electronic and structural properties, ionization potential (IP), electron affinity (EA), polarizability (α), chemical potential (μ), hardness (η), softness (S), band gap ($\Delta E_{(HOMO\text{-}LUMO)}$) and spectral characteristics of beryllium hydride oligomers by quantum chemical treatment. The discrepancies in the calculated reactive descriptors of molecules are presumably due to variations in electron distribution and spin pairing near Fermi level. The molecules $BeH_2$, $Be_3H_5$, $Be_5H_{10}$, $Be_7H_{14}$ and $Be_9H_{18}$ have been found to be more stable than $Be_2H_4$, $Be_4H_8$, $Be_6H_{12}$, $Be_8H_{16}$ and $Be_{10}H_{20}$. Ultra-soft pseudopotentials for electron-nuclei interactions with local density approximation (LDA) and generalized gradient approximation (GGA) have been used to understand the nature of bonding in crystalline beryllium hydride. The observed band gap value of crystalline $BeH_2$ is comparable with the band gap value of beryllium hydride oligomers.

***Keywords*****:** Density functional theory; stability; ionization and chemical potentials; electron affinity

[*]*Corresponding author*. Tel: +91 40 23138709; Fax: +91 40 23012800.

*E-mail*: gvsp@uohyd.ernet.in (G. Vaitheeswaran)

## 1. Introduction

The search for new potential compounds with desired properties for hydrogen storage has been a continuous challenge among researchers for past few decades [1, 2]. Hydrogen storage in the solid state is a better choice compared to the storage in gaseous or liquid state, because storage in solid state is safer, less energy intense and requires minimum space. The alkali and alkaline metal hydrides, amides, borates and alanates are known to be hydrogen storage compounds with several promising applications [3-6]. Beryllium hydride ($BeH_2$) is a simple light weight metal hydride well known for its hydrogen storage capability. The crystalline $BeH_2$ was reported as a three dimensional array of tetrahedral $BeH_4$ molecules and not as $BeH_2$ chains [7-9].

The beryllium hydride oligomers are obtained from the excited beryllium ($^1P_o$) atom and hydrogen molecule [10]. These are very difficult to synthesize and easily hydrolyzed by dilute acids. Because of its high hydrogen storage capacity (18.2 wt%), it is of considerable interest especially for use in nuclear reactor and also as a rocket fuel [5]. Therefore, understanding the geometry, stability and electronic properties of beryllium hydride oligomers is essential for developing chemical methods of hydrogen release and recovery of oligomers. Several theoretical studies emphasizing the electron correlations [11], static dipole polarizabilities [12], and electronic structures [13] have been reported elsewhere. Chen et al. [14] have reported geometry and binding energies of possible beryllium hydride clusters $(BeH_2)_n$ (where, n = 1 to 4). However, the stability of these oligomers towards hydrogen storage applications is not clearly defined. Hence, an in-depth theoretical investigation is deemed necessary to explore the various properties such as stability, reactivity, electronic structure and spectroscopic parameters of molecules.

In the present study, density functional theory calculations at HF/6-311++G**, B3LYP/6-31+G**, B3LYP/6-311++G**, B3P86/6-31+G** and B3P86/6-311++G** levels have been performed to investigate geometry, stability, reactivity, spectral parameters and electronic structure properties of model molecules. The stabilities of beryllium hydride oligomers are determined from the frontier molecular orbital energies and their gaps.

## 2. Theory and Computational detail

Ab initio molecular orbital calculations have been performed for the model molecules using the Gaussian 03 quantum chemical package [15]. Structure optimizations have been performed at the above mentioned levels. All the geometric parameters were allowed to be optimized and no constraints were imposed on the molecular structure during the optimization process. Vibrational frequencies were calculated for the optimized structures to enable us to characterize the nature of the stationary points, zero-point vibrational energy (ZPVE) and thermal correction ($H_T$). All stationary points have been positively identified as local minima with no imaginary frequencies. All the correction terms were estimated by using the set of equations [16]:

$$\left[H(T) - H(0)\right]_{trans} = \frac{5}{2} RT \tag{1}$$

$$\left[H(T) - H(0)\right]_{rot} = \frac{3}{2} RT \tag{2}$$

$$\left[H(T) - H(0)\right]_{vib} = RT \sum_{i=1}^{f} \left(\frac{h\nu_i}{kT}\right) \frac{\exp \frac{-h\nu_i}{kT}}{\left(1 - \exp \frac{-h\nu_i}{kT}\right)} \tag{3}$$

where k is the Boltzmann's constant, h is the Planck's constant, f is the number of vibrational degrees of freedom (3N-5 for linear and 3N-6 for nonlinear molecules with N being number of atoms in the molecule), $v_i$ is $i^{th}$ vibrational frequency, R is the gas constant, H is the enthalpy and T is the temperature.

Total energy has been used to evaluate reactive descriptors such as ionization potential (IP), electron affinity (EA), polarizability (α), chemical potential (μ), hardness (η), softness (S) and $\Delta E_{(HOMO-LUMO)}$ of model molecules. The chemical potential ($\mu$) and the hardness ($\eta$) of the molecules were calculated according to ref. [17,18]

$$\mu = (\partial E/\partial N)_{v(r)} \quad (4)$$

$$\eta = (\partial^2 E/\partial N^2)_{v(r)} \quad (5)$$

Electronic structure calculations based on Kohn-Sham density functional theory are more efficient and accurate and therefore the DFT version of Koopman's theorem naturally becomes an interesting topic [19]. The Koopman's theorem implemented in the density functional theory (DFT) will be less precise in contrast to HF (where the values are overestimated) because the Kohn-Sham orbital energies in DFT differ from IP by at least 2 eV which is due to the presence of electronic self interactions [20]. However, the calculated IP values of the present $BeH_2$ oligomers differ by an amount which is roughly same for all. For comparison purpose we also presented the IP and EA values with HF along with DFT values in supporting information. Koopman's theorem has been used to calculate the ionization potential and electron affinity of the beryllium hydride oligomers. The ionization potential and electron affinity of the molecule are in general equal to the negative of the highest occupied molecular orbital (HOMO) and the lowest unoccupied molecular orbital (LUMO), respectively [21, 22].

The first principle calculations of crystalline $BeH_2$ were also carried out by using the plane wave pseudopotential density functional method using CAmbridge Series of Total Energy Package (CASTEP) [23, 24]. We have used ultra-soft pseudopotentials for electron-nuclei interactions with local density approximation (LDA) and generalized gradient approximation (GGA). A plane wave basis set with energy cut-off 480 eV has been applied. For the Brillouin zone sampling, the 5 x 8 x 5 Monkhorst and Pack [25] mesh has been used.

## 3. Results and discussion

This section presents and discusses the results including the geometry, electronic properties, ionization and chemical potentials, electron affinities, softness, hardness, polarizability infrared and Raman spectra of model compounds.

### *3.1. Optimized structures*

At the outset, we have performed geometry optimizations of molecules, the optimized structures of model oligomers from the B3LYP/6-311++G** level are shown in Figure 1. The terminal and bridge Be-H lengths in all molecules are in agreement with the experimental values of ~1.33 and 1.44 ± 0.04 Å, respectively [10,11]. The calculated Be-H-Be angles vary from 84.54° to 85.91° with the method adopted and also with the size of basis set. The molecules are asymmetric and are tetragonal with four Be-H bonds i.e., there are two H-Be-H-Be-H planes that are perpendicular to each other in the $Be_3H_6$ to $B_{10}$-$H_{20}$ molecules. With increase the basis set as well as the size of the molecules the point groups have been changing from $D_{\infty h}$ to Cl. The point groups of $BeH_2$, $Be_2H_4$, $Be_3H_6$, and $Be_4H_8$ to $B_{10}H_{20}$ molecules at the three levels (HF, B3LYP and B3P86) of calculations are $D_{\infty h}$, C2v, C2 and C1, respectively. The total energies ($E_0$) and zero point vibrational energies (ZPVE) of the oligomers computed from the HF, B3LYP and B3P86 methods with basis sets 6-31+G** and 6-311++G** are presented in Table 1. The total energies of the molecules calculated from the basis set 6-311++G** with three methods happen to be nearly same. The binding energies (BE) of oligomers apparently higher, reveals that the stability of model molecules increases with size. The binding energies and lowest harmonic frequencies of molecules are summarized in Table 2.

Furthermore, we have optimized the crystalline $BeH_2$ within LDA and GGA, the optimized crystalline structure of $BeH_2$ is shown in Figure 2. The crystalline beryllium hydride contains twelve $BeH_2$ units with network of $BeH_4$ tetrahedral molecules in the unit cell. The structural parameters such as lattice constants, atomic positions and the band gap with experimental values are listed in Table 3. The calculated properties have been found to be in good agreement with the experimental values [8]. We also have calculated the electronic band structure, total and partial density of states of $BeH_2$. The density of states and band structure of the compound are shown in Figure 3. The direct band gap along Γ- Γ direction of the crystalline $BeH_2$ has been calculated to be 5.51eV with LDA and 5.58 eV with GGA. The valence band consists of Be-2p states and H-1s states and the conduction band is dominated by the Be-2s states. The overlap of Be-2p states and H-1s states at Fermi level indicates the presence of sp hybridization in the molecule with covalent character.

*3.2. Electron affinity, ionization and chemical potentials*

The ionization potential (IP) and electron affinities (EA) provide the information about the reactivity and stability of the compounds. It is well known that the compounds with higher IP and lower EA values are more stable compared to the compounds with lower IP and higher EA values. The chemical potential (μ) characterizes the tendency of electron to escape from the molecule in the equilibrium state. Koopman's theorem has been used to calculate IP, EA and μ of the molecules [21]. Electron affinity, ionization and chemical potentials of the model molecules are shown in Figure 4. It has been observed that the IP and EA values of oligomers vary with the method and the basis set. The IP values decreased with the size of the oligomers. However, the IP of $Be_2H_4$ is less compared to that of $Be_3H_6$ molecule. A maximum EA value (~0.095 a.u) for the $Be_2H_4$ has been noticed. However, we have found that the EA values of $Be_2H_4$, $Be_4H_8$, $Be_6H_{12}$, $Be_8H_{16}$ and $Be_{10}H_{20}$ molecules are higher compared to $BeH_2$, $Be_3H_5$, $Be_5H_{10}$, $Be_7H_{14}$ and $Be_9H_{18}$ molecules. The higher EA values of $Be_2H_4$, $Be_4H_8$, $Be_6H_{12}$, $Be_8H_{16}$ and $Be_{10}H_{20}$ molecules indicates that these could be less stable or highly reactive compared to $BeH_2$, $Be_3H_5$, $Be_5H_{10}$, $Be_7H_{14}$ and $Be_9H_{18}$ molecules. The variation in the EA values of the molecules is probably due to the charge transfer among the atoms within the molecule. The chemical potentials (μ) values of $Be_2H_4$, $Be_4H_8$, $Be_6H_{12}$, $Be_8H_{16}$ and $Be_{10}H_{20}$ molecules are higher than $BeH_2$, $Be_3H_5$, $Be_5H_{10}$, $Be_7H_{14}$ and $Be_9H_{18}$ molecules.

### *3.3. Hardness, softness and polarizability*

The hardness (η) and softness (S) are the important factors of the charge transfer resistance and these are inversely proportional to each other of the compound [26]. The calculated hardness and softness of model oligomers are presented in Figure 5. It is found that the hardness and softness values are independent of the method and choice of basis set. According to the principle of maximum hardness (PMH), the molecule in the equilibrium state possess maximum hardness and in the transition state posses minimum hardness [27, 28]. It has been found that the $BeH_2$, $Be_3H_5$, $Be_5H_{10}$, $Be_7H_{14}$ and $Be_9H_{18}$ molecules are more stable than others. The dipole plarizability (α) measures the distortion of the electron density of the molecules and the response of the molecule to the external electric field. The polarizability increases rapidly with size of the oligomers i.e. distortion of the electronic density of oligomer is linear with external electric field and it is in good agreement with the previous studies [13].

### *3.4. Frontier molecular orbital energies*

The stability of the compounds is usually measured by the total energy (for isomers), bond length, bond dissociation energy, frontier orbital energy and their gaps, and so on. We have used frontier orbital energies to evaluate stability. Fukui et al [29] noticed the prominent role played by highest occupied molecular orbital (HOMO) and lowest unoccupied molecular orbital (LUMO) in governing chemical reactions. It has been revealed in several investigations that the band gap ($\Delta E_{HOMO-\ LUMO}$) in energy between the HOMO and LUMO is an important stability index of the molecules [30-33]. A large band gap implies high stability and small gap implies low stability. The high stability in turn indicates low chemical reactivity and small band gap indicates high chemical reactivity. In other words, smaller the band gap between HOMO and LUMO, easier the electron transition and lesser the stability of the compound will be. The frontier molecular orbitals of model molecules obtained from the B3LYP/6-311++G** level are shown in Figure 6. With increase of the monomer units $BeH_2$ of the oligomer, the HOMO has been observed to be a finite graphene like structure [34] and observed to be prominent in the model molecules from $Be_4H_8$ to $Be_{10}H_{20}$. As with increase of the basis set, the frontier molecular orbital energies and their gaps have been increased. The band gap values have been decreased from $BeH_2$ to $B_{10}H_{20}$ with the method and the basis set. The band gaps of molecules vary from 0.294 to 0.242 a.u with the level of calculation and also with size of the molecule. The model compounds $BeH_2$ is most stable and $Be_{10}H_{20}$ is least stable reflected from the band gap values. The frontier molecular orbital energies of $BeH_2$, $Be_3H_5$, $Be_5H_{10}$, $Be_7H_{14}$ and $Be_9H_{18}$ have higher values compared to $Be_2H_4$, $Be_4H_8$, $Be_6H_{12}$, $Be_8H_{16}$ and $Be_{10}H_{20}$ molecules. The frontier molecular orbital energies of the model molecules are shown in Figure 7. $BeH_2$ oligomers structures are similar like zig zag edge states of finite graphene [34]. The edge states in non metallic graphene ribbons are located on length confined zig zag edge as discussed by Shemella et al [34], we also observe a similar zig zag edge in the present oligomers. Therefore one can expect the similar looking of HOMOs in $BeH_2$ oligomers like finite graphene or nanotubes where they are located at the edges. In finite graphene ribbons this typical behavior is observed only for non metallic ribbons and which is not the case for metallic graphene ribbons. In the present case we observe the HOMO energy levels located at the length confined edge of the oligomers. Since HOMOs are localized at the edges, it may be possible to control the band gaps using different functionalized groups.

### *3.5. Mulliken atomic charges*

One of the most important property of a molecule is its charge or spin density distribution. Although there is no unique definition of how many electrons are attached to an atom in a molecule it has nevertheless proven useful in many cases to perform such population analysis. Due to its simplicity the Mulliken population analysis has become the most familiar method to count electrons to be associated with an atom. It has been found that the hydrogen and beryllium atoms have negative and positive Mulliken atomic charges, respectively. However, beryllium atoms in $Be_4H_8$, $Be_5H_{10}$ and $Be_6H_{12}$ molecule have negative charges of two, one and four, respectively. We have noticed only two negative charges in $Be_7H_{14}$, $Be_8H_{16}$, $Be_9H_{18}$ and $Be_{10}H_{20}$ molecules. In general, if more electrons are shared by the hydrogen atom, higher is the negative charge. The higher negative charges have been observed for the beryllium atoms in $Be_4H_8$ to $Be_{10}H_{20}$ molecules and thus there is a possibility of covalent bonding among beryllium and hydrogen atoms.

The Mulliken population on the charge density distribution of crystalline $BeH_2$ is shown in Figure 8. The charges of beryllium and hydrogen atoms of crystalline $BeH_2$ have been calculated to be -0.23, -0.24, 0.50 and 0.45 e for H(1), H(2), Be(1) and Be(2) atoms, respectively. The negative charge of hydrogen atoms implies the electronegative nature towards beryllium atoms. R. S. Mulliken [35] states that "If the total overlap population between two atoms is positive, they are bonded; if negative, they are antibonded". The high positive population indicates the possibility of having high degree of covalent character of bond [35, 36]. All Be-H bonds have bond population about 0.55 indicating the equal ionic and covalent character. Figure 8 shows the charge density distribution of crystalline $BeH_2$. The charge density around H atom has been found to be higher compared to beryllium atom. The shape of charge density around hydrogen atoms has been appearing to be slightly deformed towards beryllium atoms. Therefore, there may be possibility of equal covalent and ionic character of Be-H bonds in $BeH_2$.

### *3.6. Infrared and Raman spectra*

The infrared (IR) and Raman spectra of molecules are obtained from the computed structures at B3LYP/6-31++G** levels are presented in Figure 9 & 10. The calculated wave numbers of $BeH_2$ and $Be_2H_4$ molecules are in good agreement with experimental values [7-9]. It was observed that the stretching vibration (1510 $cm^{-1}$) of $Be_3H_6$ molecule has been found to be splitting and has been found to be continued in all oligomers. The splitting of vibrational line of 1510 $cm^{-1}$ has been observed in $Be_3H_6$ to $Be_{10}H_{20}$ molecules. However, the wave numbers corresponding to 1900, 1910, 1920 and 1930 $cm^{-1}$ in Raman spectra of $Be_{10}H_{20}$ probably due to the splitting of the line corresponding to 1810 $cm^{-1}$.

## 4. Conclusion

The electronic structure properties of beryllium hydride oligomers have been carried out at HF, B3LYP and B3P86 methods with 6-31+G** and 6-311++G** basis sets. The band gap $\Delta E_{(LUMO-HOMO)}$, ionization potential (IP), electron affinity (EA), chemical potential ($\mu$), hardness ($\eta$), softness (S) and polarizabilities ($\alpha$) of beryllium hydride oligomers have been calculated. From the band gap, chemical potential, hardness and softness of molecules, $BeH_2$, $Be_3H_5$, $Be_5H_{10}$, $Be_7H_{14}$ and $Be_9H_{18}$ molecules have been found to be more stable compared to $Be_2H_4$, $Be_4H_8$, $Be_6H_{12}$, $Be_8H_{16}$ and $Be_{10}H_{20}$ molecules. The IR spectra of $BeH_2$ and $Be_2H_4$ molecules were found to be in good agreement with the experimental values. The IR spectra of molecules have five fundamental bands (475 to 575, 840 to 890, 1410 to 1520, 1610 and 2150 $cm^{-1}$) that are prominent.

## Acknowledgements

Ch B L and K R B acknowledge the sustaining financial support from CAS-RFSMS and DRDO through ACRHEM, University of Hyderabad. Authors thank CMSD, University of Hyderabad for computational time.

# Figure Captions

**Figure 1**. (Color online) Optimized structures of model molecules computed at the B3LYP/6-311++ G** level. The bond lengths are in angstroms and bond angles are in degrees.

**Figure 2**. (Color online) Optimized crystalline structure of $BeH_2$ with local density approximation (LDA).

**Figure 3.** (Color online) The density of states and band structure of crystalline $BeH_2$ with local density approximation (LDA).

**Figure 4.** (Color online) Electron affinity, ionization and chemical potentials of the model molecules computed from the various levels of theory.

**Figure 5.** (Color online) The calculated hardness and softness of model oligomers computed from the various levels of theory.

**Figure 6.** (Color online) Frontier molecular orbitals of the model molecules computed from the B3LYP/6-311++G** level.

**Figure 7.** (Color online) The HOMO-LUMO gap of the model molecules computed from the B3LYP/6-311++G** level.

**Figure 8.** (Color online) The charge density distribution of the crystalline $BeH_2$ with local density approximation (LDA).

**Figure 9.** (Color online) The infrared (IR) spectra of the molecule are obtained from the computed structures from the B3LYP/6-31++G** level.

**Figure 10.** (Color online) Raman spectra of the molecule are obtained from the computed structures from the B3LYP/6-31++G** level.

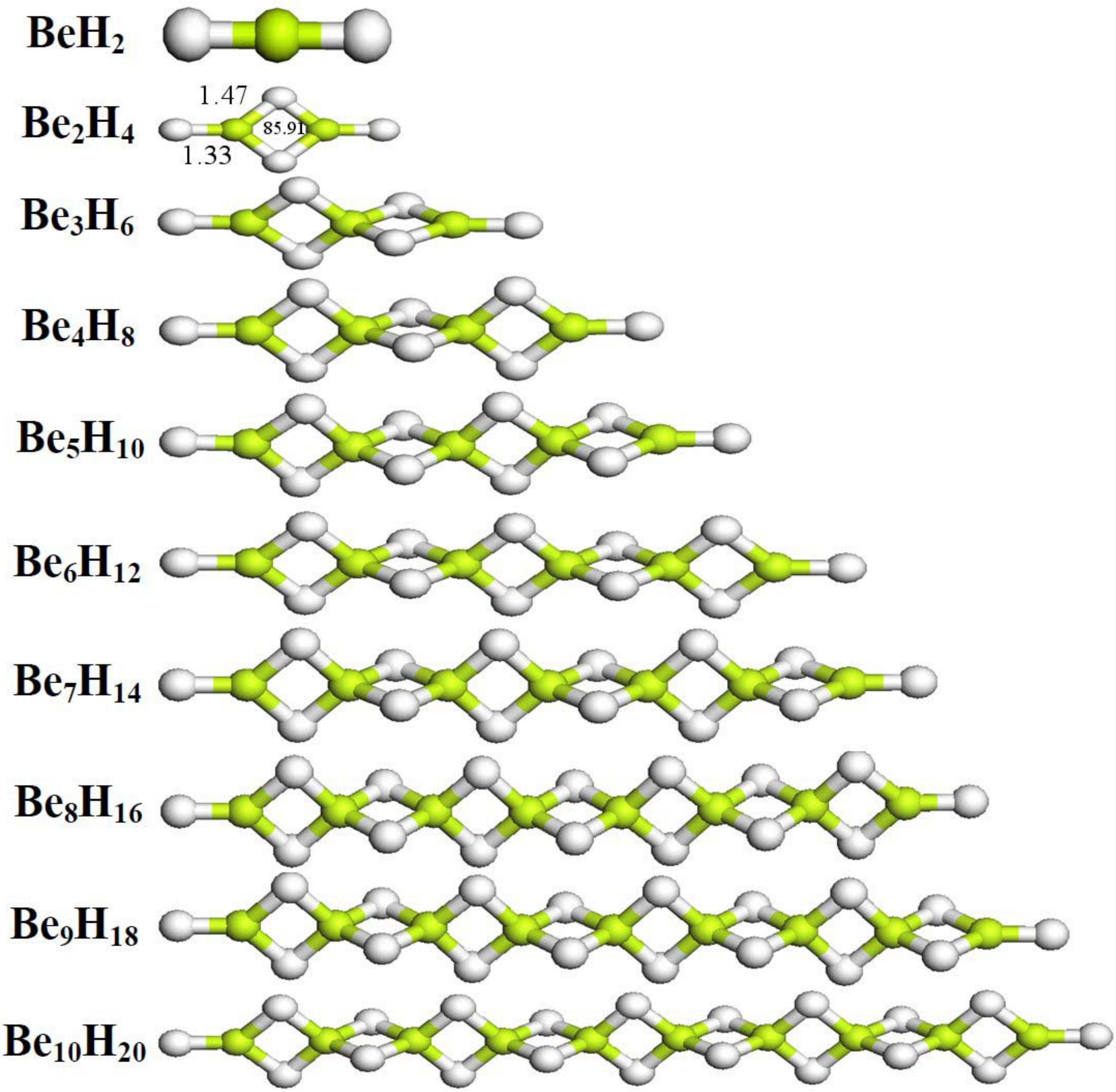

**Fig 1**. Bheem et al

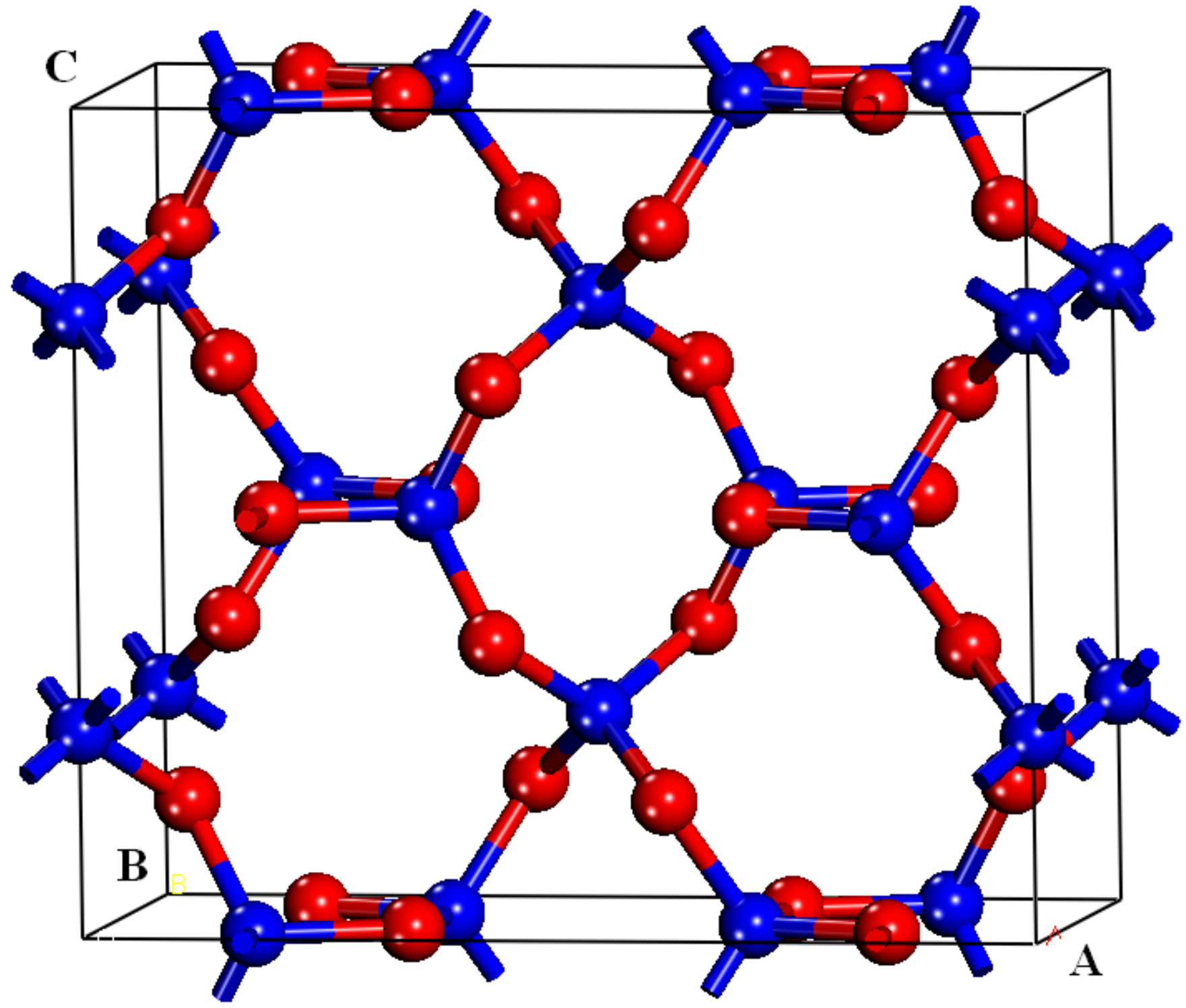

**Fig 2**.  Bheem et al

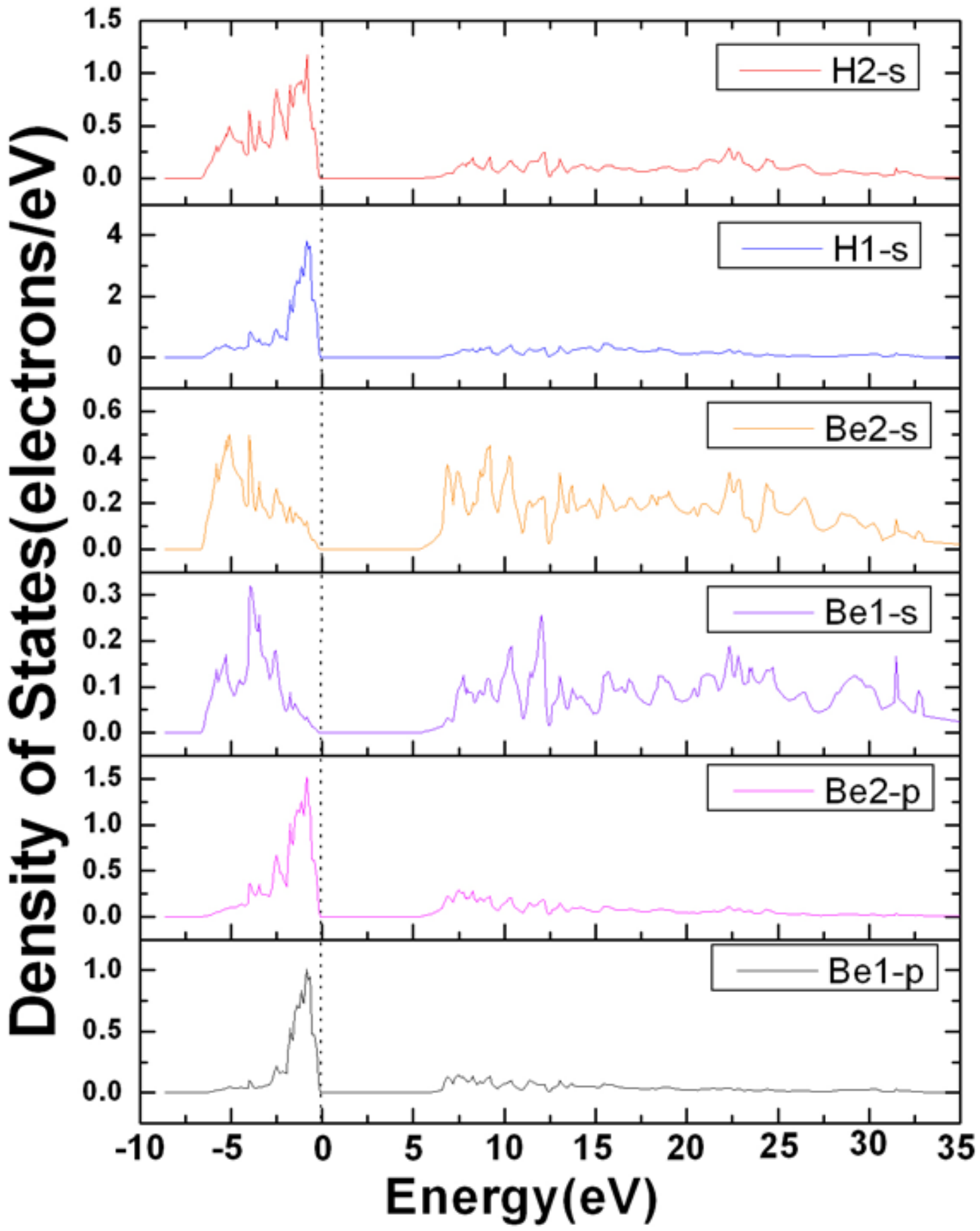

(a)

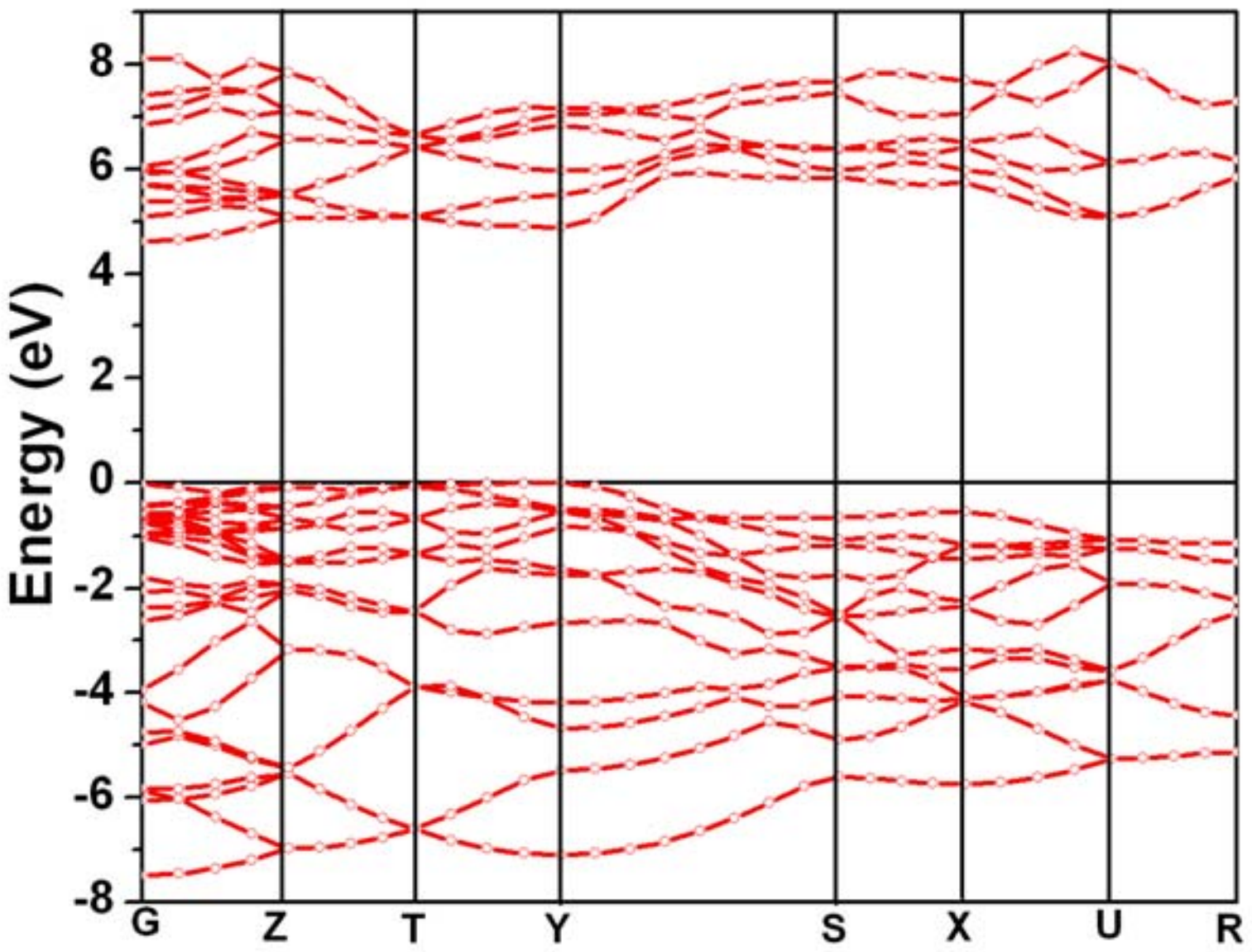

**Fig 3**. Bheem et al

(b)

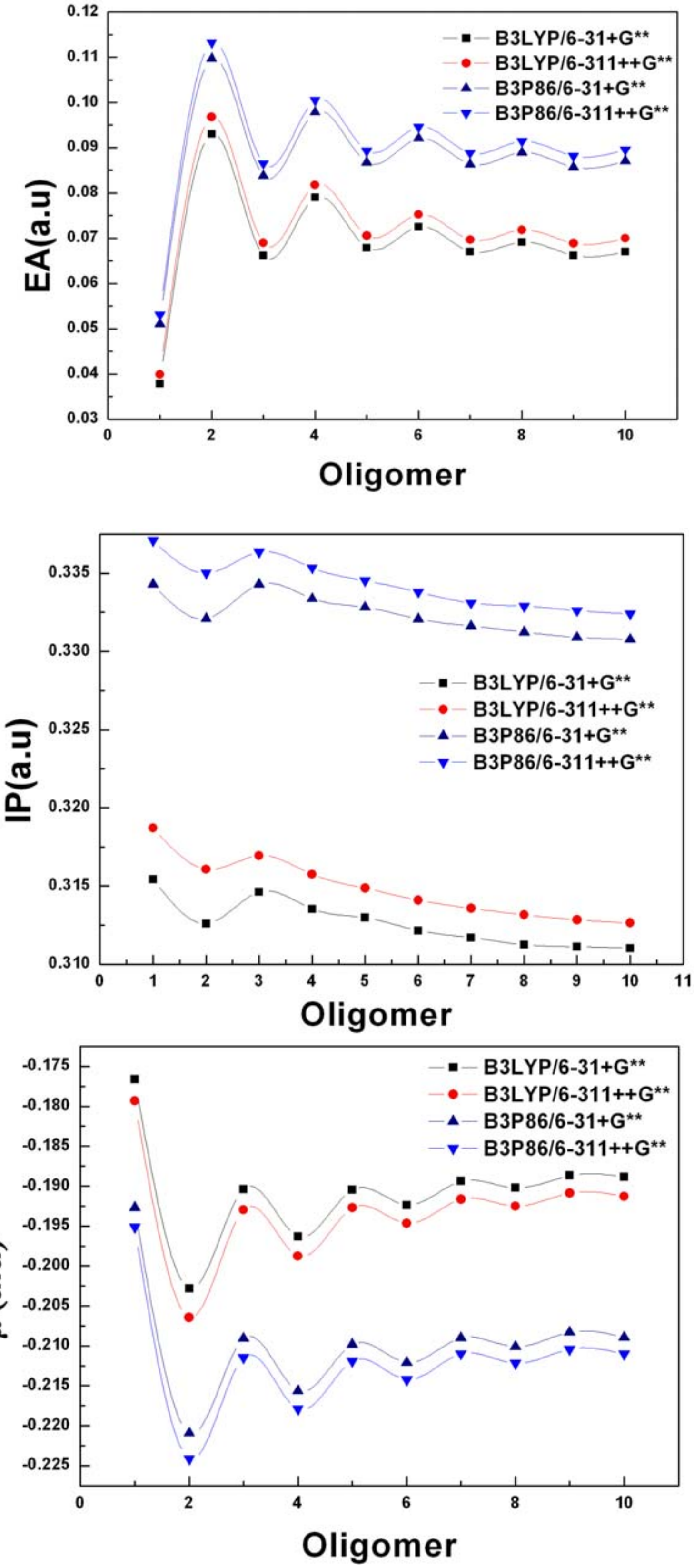

(a)

(b)

(c)

**Fig 4**. Bheem et al

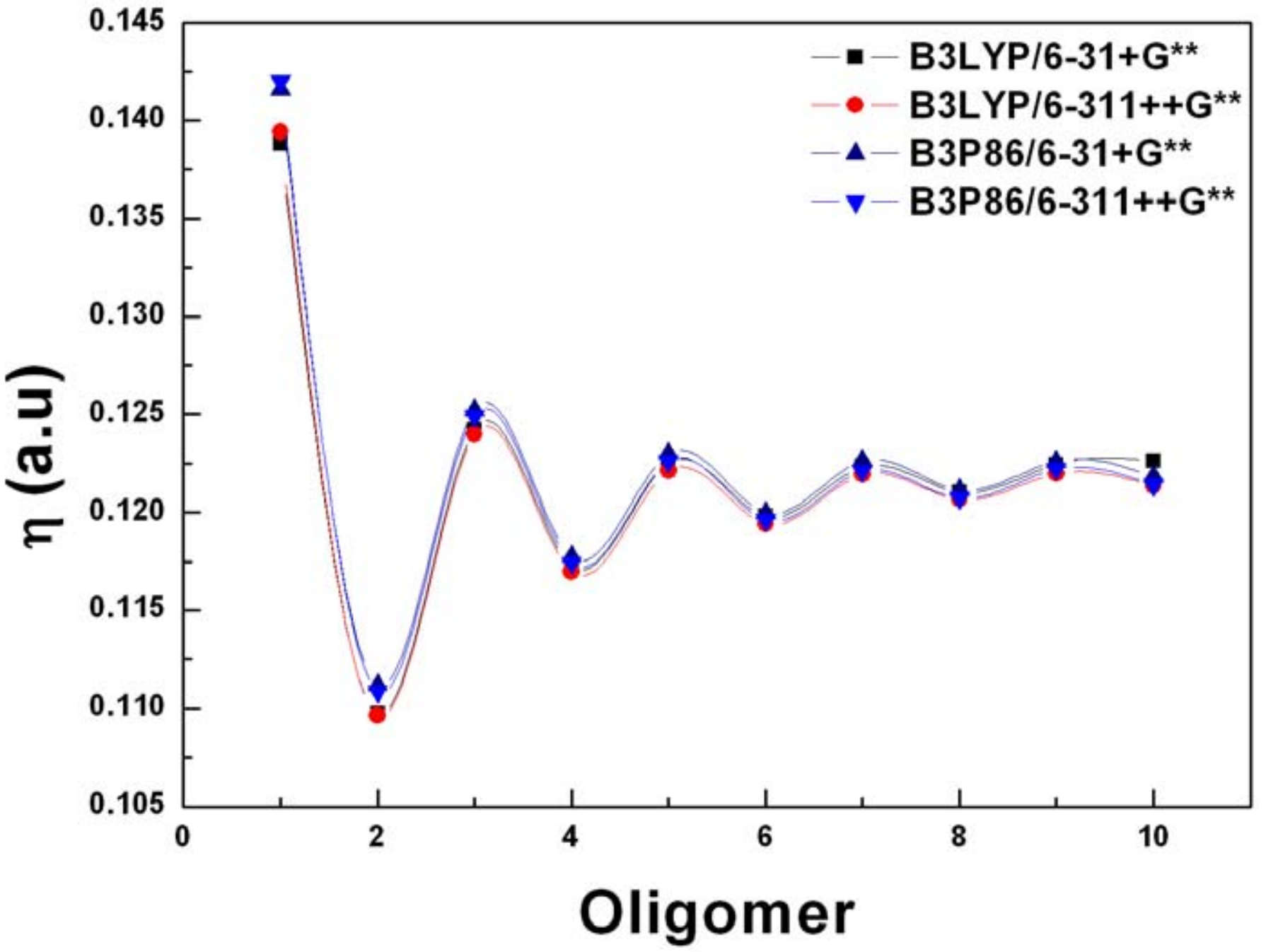

(a)

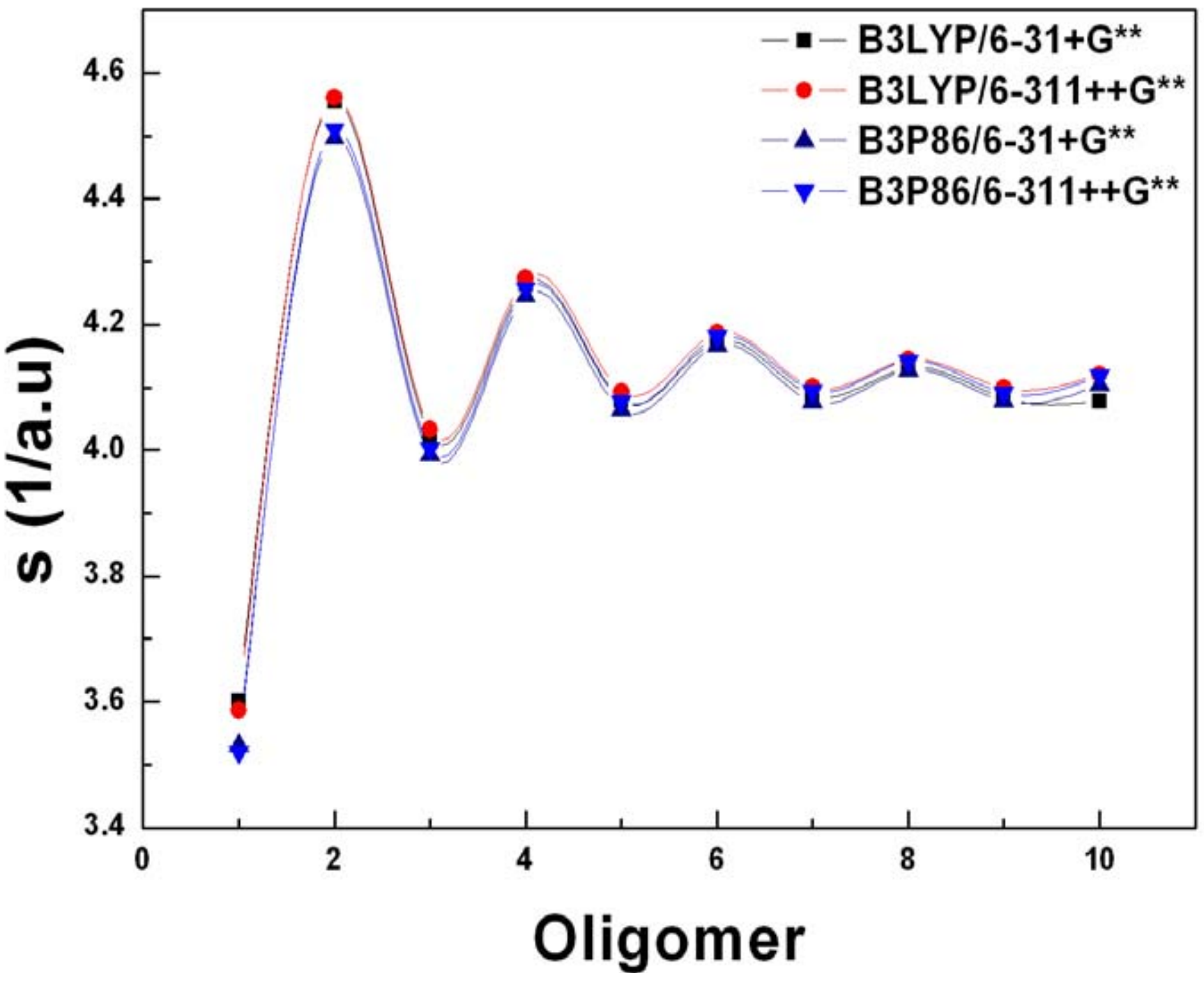

(b)

**Fig 5** Bheem et al

| Oligomer | HOMO | LUMO |
|---|---|---|
| $BeH_2$ | 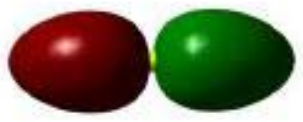 | 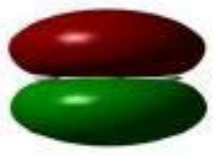 |
| $Be_2H_4$ | 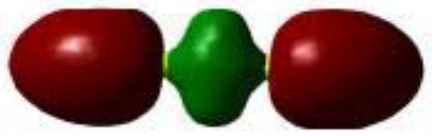 | 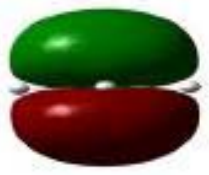 |
| $Be_3H_6$ | 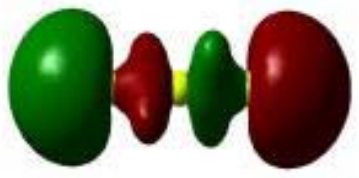 | 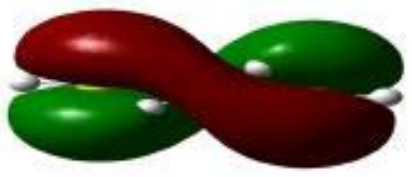 |
| $Be_4H_8$ | 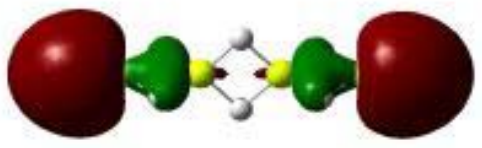 | 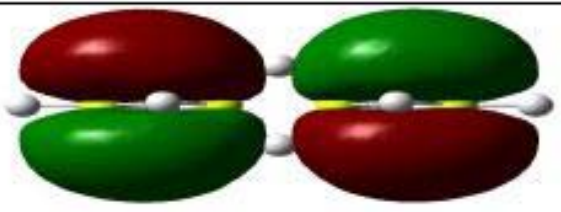 |
| $Be_5H_{10}$ | 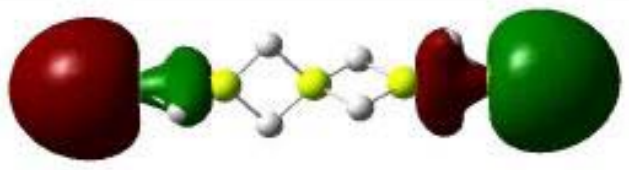 | 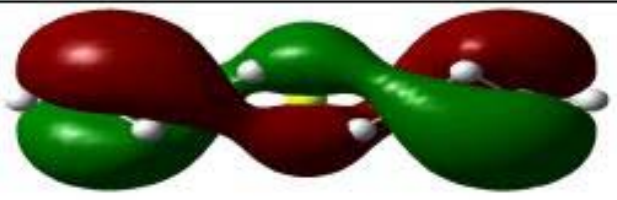 |
| $Be_6H_{12}$ | 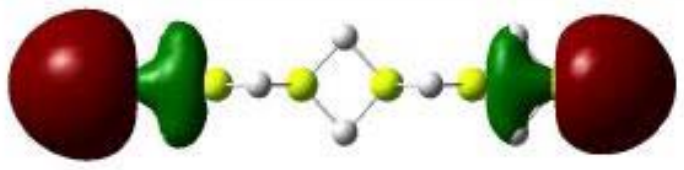 | 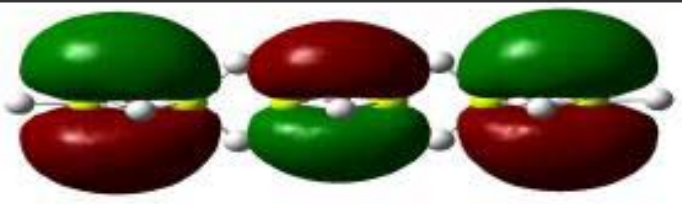 |
| $Be_7H_{14}$ | 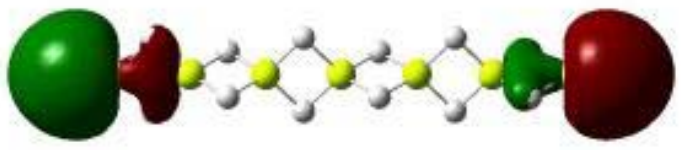 | 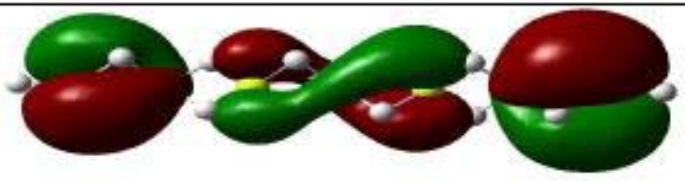 |
| $Be_8H_{16}$ | 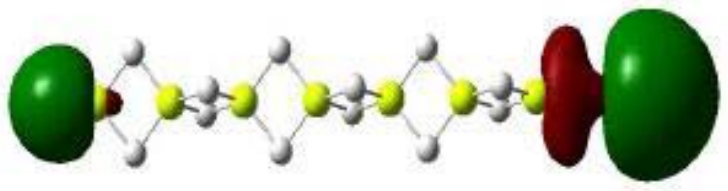 | 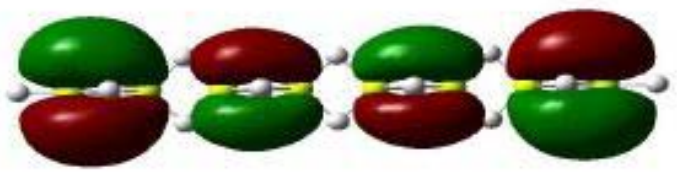 |
| $Be_9H_{18}$ | 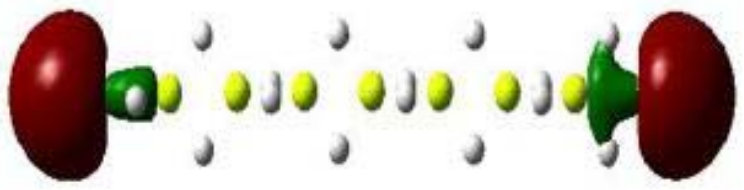 | 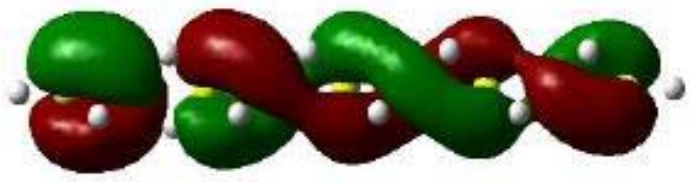 |
| $Be_{10}H_{20}$ | 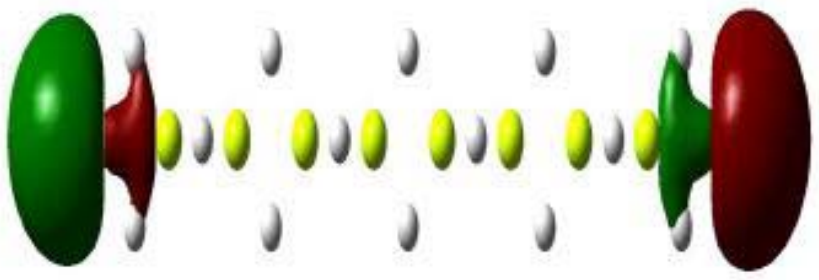 | 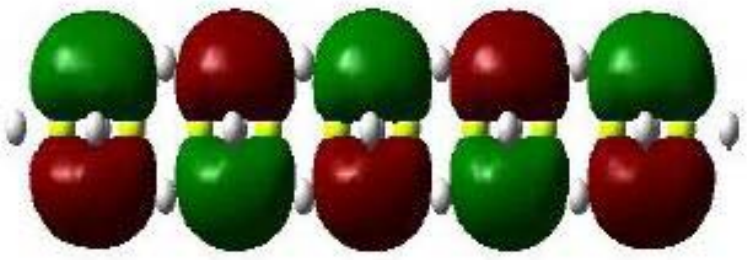 |

**Fig 6.** Bheem et al

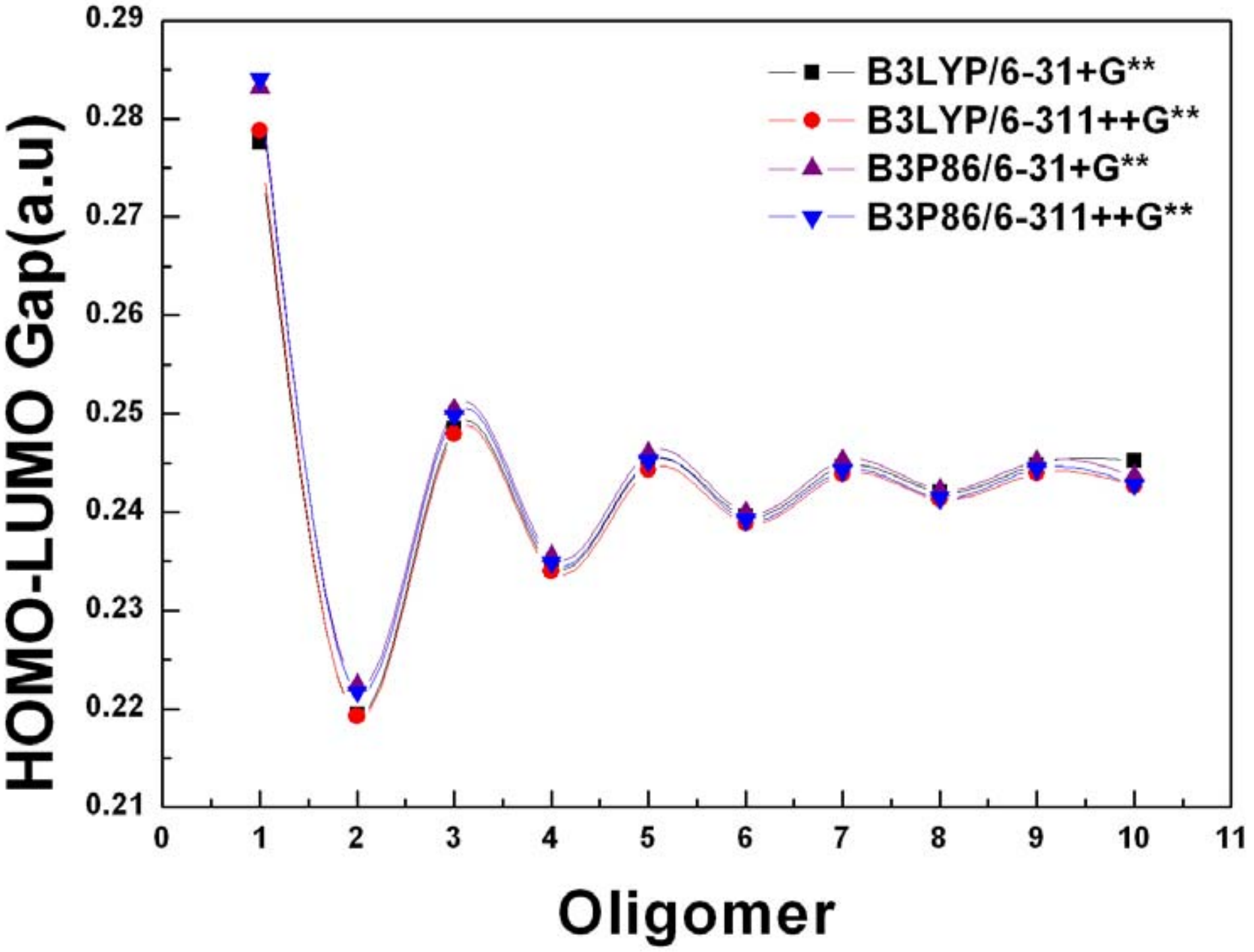

**Fig 7.** Bheem et al

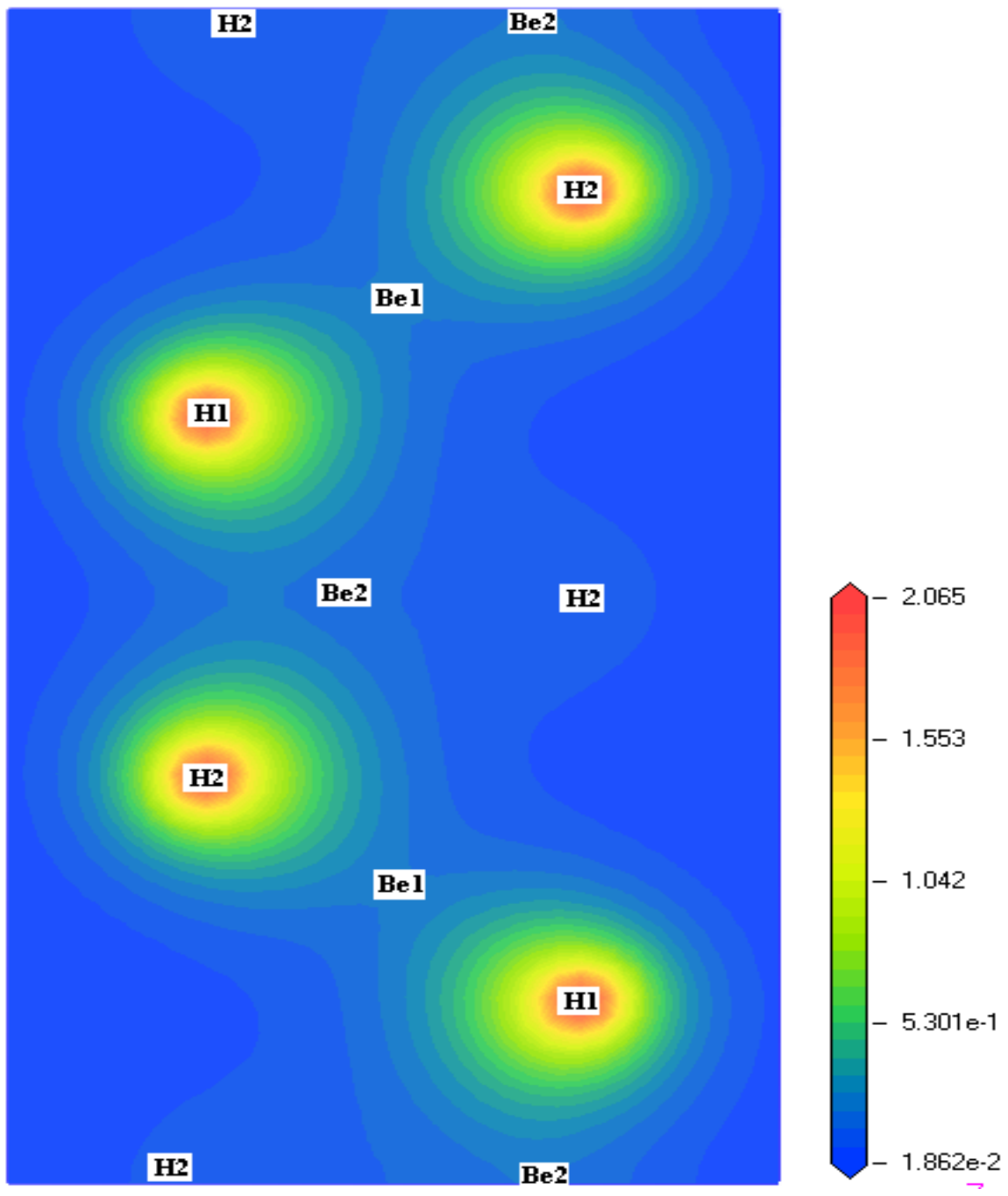

**Fig 8.** Bheem et al

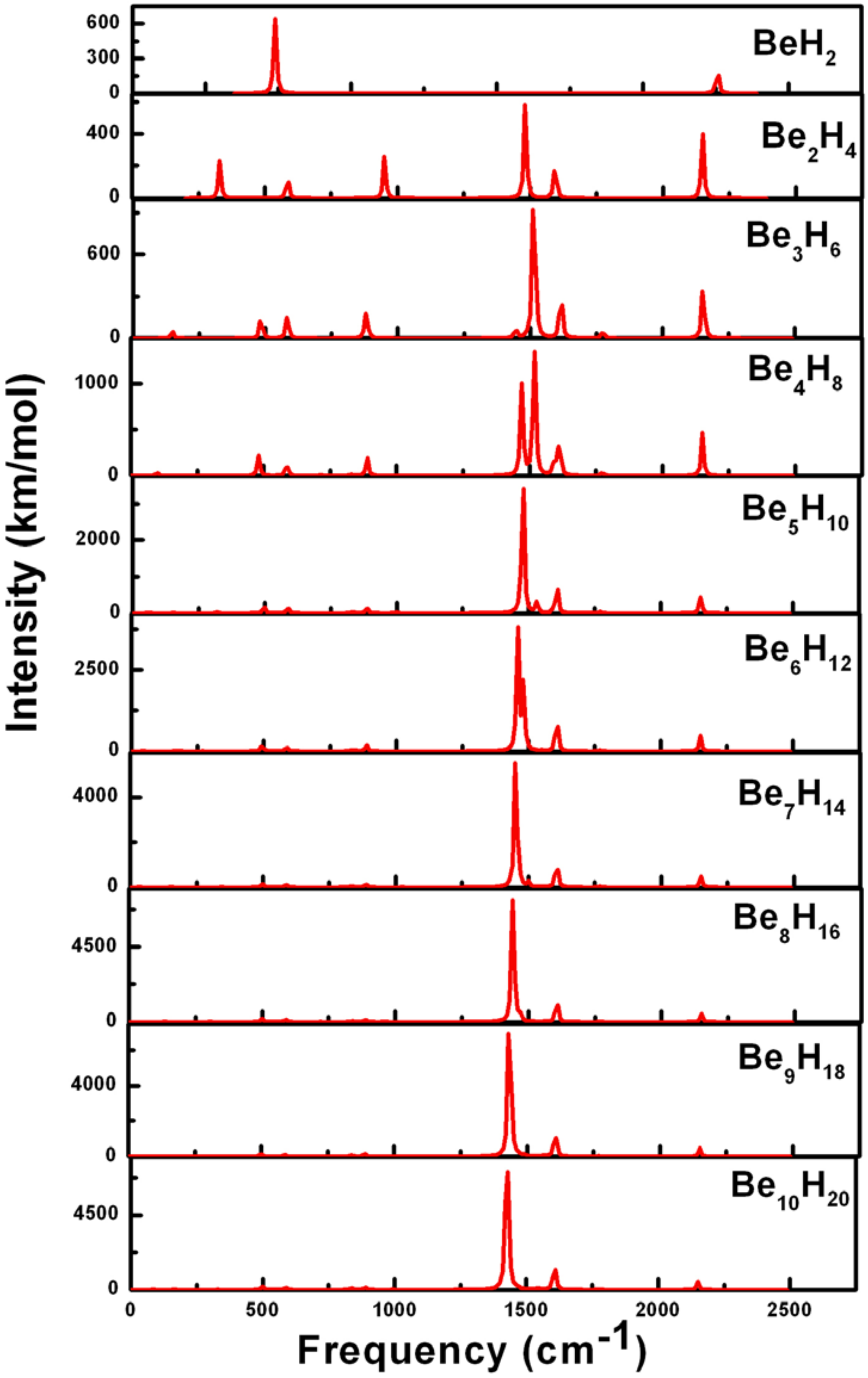

**Fig 9.** Bheem et al

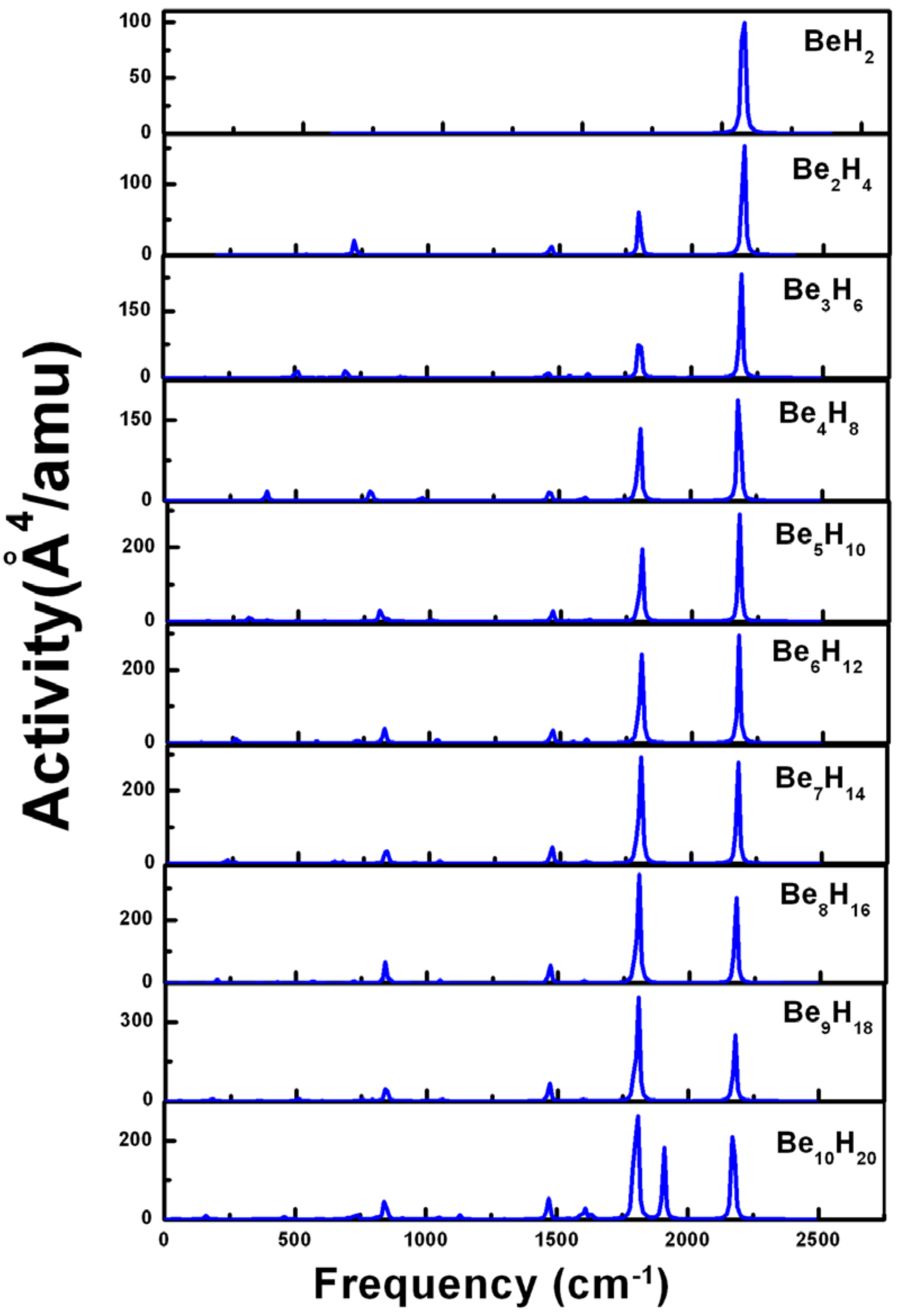

**Fig 10.** Bheem et al

**Table 1.** The total energy ($E_0$), thermal correction to enthalpy ($H_T$) and zero-point vibrational energy (ZPVE) of oligomers[a] computed from the various levels of theory.

| Compd | HF | | | B3LYP | | | | | | B3P86 | | | | | |
|---|---|---|---|---|---|---|---|---|---|---|---|---|---|---|---|
| | 6-311++G** | | | 6-31+G** | | | 6-311++G** | | | 6-31+G** | | | 6-311++G** | | |
| | $E_0$ | $H_T$ | ZPVE | $E_0$ | $H_T$ | ZPVE | $E_0$ | $H_T$ | ZPVE | $E_0$ | $H_T$ | ZPVE | $E_0$ | $H_T$ | ZPVE |
| $BeH_2$ | -15.770655 | 0.01684 | 8.37519 | -15.9194642 | 0.016651 | 8.25015 | -15.9227871 | 0.01665 | 8.2547 | -16.014507 | 0.01656 | 8.18875 | -16.017452 | 0.01655 | 8.18105 |
| $Be_2H_4$ | -31.580371 | 0.03796 | 20.79412 | -31.8891637 | 0.037273 | 20.30353 | -31.8968005 | 0.03739 | 20.39367 | -32.085009 | 0.03717 | 20.2243 | -32.091615 | 0.03726 | 20.29887 |
| $Be_3H_6$ | -47.400868 | 0.06020 | 33.80599 | -47.8732126 | 0.059351 | 33.25028 | -47.8833864 | 0.05938 | 33.23046 | -48.170142 | 0.05922 | 33.12694 | -48.178978 | 0.05920 | 33.07756 |
| $Be_4H_8$ | -63.219947 | 0.08225 | 46.6076 | -63.8543623 | 0.081193 | 45.94943 | -63.8677149 | 0.08118 | 45.87414 | -64.252376 | 0.08101 | 45.77318 | -64.263967 | 0.08094 | 45.66356 |
| $Be_5H_{10}$ | -79.039124 | 0.10430 | 59.35205 | -79.8355828 | 0.102975 | 58.56376 | -79.8522783 | 0.10294 | 58.4514 | -80.334774 | 0.10274 | 58.35945 | -80.349236 | 0.10265 | 58.19407 |
| $Be_6H_{12}$ | -94.858246 | 0.12629 | 72.02403 | -95.8166281 | 0.124795 | 71.19757 | -95.8367255 | 0.12471 | 70.99464 | -96.416991 | 0.12453 | 70.94126 | -96.434395 | 0.12422 | 70.5179 |
| $Be_7H_{14}$ | -110.67735 | 0.14843 | 84.88245 | -111.797587 | 0.146867 | 83.64777 | -111.821247 | 0.14647 | 83.68564 | -112.49917 | 0.14640 | 83.54583 | -112.51957 | 0.14605 | 83.19911 |
| $Be_8H_{16}$ | -126.49644 | 0.17046 | 97.59077 | -127.778541 | 0.168186 | 96.18721 | -127.805642 | 0.16832 | 96.19184 | -128.58130 | 0.16793 | 95.92162 | -128.60472 | 0.16788 | 95.80352 |
| $Be_9H_{18}$ | -142.31553 | 0.19255 | 110.38 | -143.759471 | 0.189793 | 108.5679 | -143.790024 | 0.18997 | 108.60116 | -144.66343 | 0.18955 | 108.2936 | -144.68982 | 0.18941 | 108.10944 |
| $Be_{10}H_{20}$ | -158.09884 | 0.21220 | 121.7454 | -159.695738 | 0.210878 | 120.5257 | -159.774477 | 0.21179 | 121.21814 | -160.74555 | 0.21109 | 120.58449 | -160.77498 | 0.21125 | 120.73607 |

[a] The total energy and thermal correction in atomic units and zero point vibrational energy in kcal/mol.

**Table 2.** The calculated binding energy (BE) and lowest harmonic frequencies ($\omega_L$)[b] of oligomers computed from the various levels of theory.

| Compd | HF | | B3LYP | | | | B3P86 | | | |
|---|---|---|---|---|---|---|---|---|---|---|
| | 6-311++G** | | 6-31+G** | | 6-311++G** | | 6-31+G** | | 6-311++G** | |
| | BE | Lowest harmonic frequency | BE | Lowest harmonic frequency | BE | Lowest harmonic frequency | BE | Lowest harmonic frequency | BE | Lowest harmonic frequency |
| **$BeH_2$** | -15.770655 | 750.4270 | -15.9194642 | 736.1061 | -15.9227871 | 739.1867 | -16.0145076 | 722.0296 | -16.0174524 | 723.6695 |
| **$Be_2H_4$** | -15.790185 | 344.5295 | -15.9445818 | 325.6742 | -15.9484002 | 326.4559 | -16.0425045 | 317.4523 | -16.0458076 | 323.1058 |
| **$Be_3H_6$** | -15.800289 | 159.2581 | -15.9577375 | 162.2520 | -15.9611288 | 147.8768 | -16.0567140 | 154.4502 | -16.059659 | 141.7508 |
| **$Be_4H_8$** | -15.804986 | 89.9369 | -15.9635905 | 97.9936 | -15.9669287 | 88.4270 | -16.0630940 | 93.7718 | -16.0659917 | 85.1131 |
| **$Be_5H_{10}$** | -15.807824 | 64.5284 | -15.9671165 | 67.3223 | -15.9704556 | 61.6708 | -16.0669549 | 67.8228 | -16.0698473 | 58.6486 |
| **$Be_6H_{12}$** | -15.809707 | 45.7854 | -15.9694380 | 48.4272 | -15.9727800 | 45.4036 | -16.0694980 | 45.9539 | -16.0723990 | 41.5977 |
| **$Be_7H_{14}$** | -15.811050 | 34.3550 | -15.9710840 | 35.2540 | -15.9744600 | 34.7928 | -16.0713112 | 32.4992 | -16.0742200 | 33.2763 |
| **$Be_8H_{16}$** | -15.812050 | 26.4227 | -15.9723100 | 26.7342 | -15.9757052 | 26.3839 | -16.0726635 | 24.3892 | -16.0755907 | 25.3722 |
| **$Be_9H_{18}$** | -15.812837 | 21.1353 | -15.9732740 | 21.6588 | -15.9766690 | 20.2304 | -16.0737146 | 18.5744 | -16.0766474 | 18.4816 |
| **$Be_{10}H_{20}$** | -15.809884 | 5.13847 | -15.9695738 | 15.7484 | -15.9774477 | 17.1295 | -16.0745554 | 15.0369 | -16.0774985 | 16.5513 |

[b] The binding energies in atomic units and lowest harmonic frequencies in $cm^{-1}$.

**Table 3.** Lattice constants, atomic positions and band gap of the crystalline $BeH_2$ along with experimental values.

| Parameter | LDA | GGA | Experiment[9] |
|---|---|---|---|
| **a(Å)** | 8.700 | 8.964 | 9.082 |
| **b(Å)** | 3.939 | 4.147 | 4.160 |
| **c(Å)** | 7.477 | 7.633 | 7.707 |
| **Atomic positions** | Be1 (0, 0, 0.25) | (0, 0, 0.25) | (0, 0, 0.25) |
| | Be2 (0.1680, 0.1217, 0) | (0.167, 0.118, 0) | (0.1699, 0.1253, 0) |
| | H1 (0.0893, 0.2481, 0.1549) | (0.088, 0.225, 0.152 ) | (0.0895, 0.1949, 0.1515) |
| | H2 (0.3219, 0.2568, 0) | (0.310, 0.274, 0) | (0.3055, 0.2823, 0) |
| **$E_g$(eV)** | 5.51 | 5.58 | - |

# Supporting Information

# Quantum chemical studies on beryllium hydride oligomers

**Ch. Bheema Lingam,**[a] **K. Ramesh Babu,**[b] **Surya P. Tewari**[a,b]**, G. Vaitheeswaran**,[b,*]

[a]*School of Physics, University of Hyderabad, Hyderabad-500046, India*

[b]*Advanced Centre of Research in High Energy Materials (ACRHEM), University of Hyderabad, Hyderabad-500046, India.*

**Table 1.** The chemical potential (μ), ionization potential (IP) and electron affinities (EA)[a] of molecules computed at the various levels of theory.

| compd | HF | | | B3LYP | | | | | | B3P86 | | | | | |
|---|---|---|---|---|---|---|---|---|---|---|---|---|---|---|---|
| | 6-311++G** | | | 6-31+G** | | | 6-311++G** | | | 6-31+G** | | | 6-311++G** | | |
| | μ | IP | EA | μ | IP | EA | μ | IP | EA | μ | IP | EA | μ | IP | EA |
| $BeH_2$ | -0.20753 | 0.44838 | -0.03332 | -0.17663 | 0.31544 | 0.03782 | -0.1793 | 0.31871 | 0.0399 | -0.19269 | 0.33429 | 0.05108 | -0.19509 | 0.3371 | 0.05308 |
| $Be_2H_4$ | -0.21176 | 0.4456 | -0.00808 | -0.20283 | 0.3126 | 0.09307 | -0.20643 | 0.31607 | 0.09679 | -0.2209 | 0.3321 | 0.1097 | -0.22414 | 0.33501 | 0.11327 |
| $Be_3H_6$ | -0.21024 | 0.44758 | -0.02109 | -0.19039 | 0.31462 | 0.06615 | -0.19296 | 0.31694 | 0.06898 | -0.20906 | 0.33429 | 0.08383 | -0.21144 | 0.33636 | 0.08653 |
| $Be_4H_8$ | -0.21057 | 0.44703 | -0.01659 | -0.19628 | 0.31354 | 0.07901 | -0.19875 | 0.31575 | 0.08176 | -0.21563 | 0.33338 | 0.09787 | -0.21791 | 0.33535 | 0.10047 |
| $Be_5H_{10}$ | -0.21007 | 0.44633 | -0.0202 | -0.19043 | 0.31298 | 0.06789 | -0.19271 | 0.31486 | 0.07056 | -0.2098 | 0.33283 | 0.08678 | -0.21191 | 0.33452 | 0.08929 |
| $Be_6H_{12}$ | -0.21047 | 0.44569 | -0.01875 | -0.19236 | 0.31216 | 0.07256 | -0.19467 | 0.31409 | 0.07524 | -0.21209 | 0.33208 | 0.0921 | -0.21421 | 0.33381 | 0.09461 |
| $Be_7H_{14}$ | -0.21033 | 0.4452 | -0.02153 | -0.18937 | 0.3117 | 0.06704 | -0.19164 | 0.31358 | 0.06969 | -0.20899 | 0.33163 | 0.08635 | -0.21098 | 0.3331 | 0.08886 |
| $Be_8H_{16}$ | -0.21047 | 0.44484 | -0.02039 | -0.19022 | 0.31125 | 0.06918 | -0.1925 | 0.31315 | 0.07184 | -0.2101 | 0.33124 | 0.08895 | -0.21217 | 0.3329 | 0.09144 |
| $Be_9H_{18}$ | -0.21006 | 0.44456 | -0.02144 | -0.18867 | 0.31112 | 0.06622 | -0.19087 | 0.31284 | 0.06889 | -0.2083 | 0.3309 | 0.08569 | -0.21041 | 0.33262 | 0.08819 |
| $Be_{10}H_{20}$ | -0.20999 | 0.44421 | -0.01962 | -0.18881 | 0.31102 | 0.0712 | -0.1913 | 0.31264 | 0.06996 | -0.20893 | 0.33077 | 0.08709 | -0.211 | 0.33241 | 0.08959 |

[a] The chemical potential (μ), ionization potential (IP) and electron affinities (EA) are in atomic units.

Table 2. The hardness (η), softness (S) and polarizability (α)[b] of molecules computed at the various levels of theory.

| compd | HF | | | B3LYP | | | | | | B3P86 | | | | | |
|---|---|---|---|---|---|---|---|---|---|---|---|---|---|---|---|
| | 6-311++G** | | | 6-31+G** | | | 6-311++G** | | | 6-31+G** | | | 6-311++G** | | |
| | η | s | α | η | s | α | η | s | α | η | s | α | η | s | α |
| $BeH_2$ | 0.24085 | 2.07598 | 14.932 | 0.13881 | 3.60205 | 15.796 | 0.13941 | 3.58667 | 16.727 | 0.14161 | 3.53095 | 15.796 | 0.14201 | 3.52088 | 16.814 |
| $Be_2H_4$ | 0.23084 | 2.15042 | 32.034 | 0.10977 | 4.55519 | 31.348 | 0.10964 | 4.56038 | 34.092 | 0.1112 | 4.4964 | 31.205 | 0.11087 | 4.50979 | 34.034 |
| $Be_3H_6$ | 0.23733 | 2.10673 | 40.238 | 0.12424 | 4.02463 | 39.73 | 0.12398 | 4.03291 | 42.815 | 0.12523 | 3.99265 | 39.574 | 0.12492 | 4.00272 | 42.829 |
| $Be_4H_8$ | 0.23447 | 2.12315 | 48.38 | 0.11726 | 4.26385 | 47.45 | 0.117 | 4.27369 | 51.04 | 0.11776 | 4.2461 | 47.326 | 0.11744 | 4.25749 | 51.552 |
| $Be_5H_{10}$ | 0.23577 | 2.11576 | 62.899 | 0.12255 | 4.08013 | 61.089 | 0.12215 | 4.09333 | 61.767 | 0.12303 | 4.06421 | 61.952 | 0.12262 | 4.0778 | 66.16 |
| $Be_6H_{12}$ | 0.23422 | 2.12567 | 141.178 | 0.1198 | 4.17362 | 159.784 | 0.11943 | 4.18673 | 159.497 | 0.11999 | 4.16701 | 160.909 | 0.1196 | 4.1806 | 159.183 |
| $Be_7H_{14}$ | 0.23486 | 2.12488 | 168.241 | 0.12233 | 4.0873 | 191.91 | 0.12195 | 4.10021 | 193.545 | 0.12264 | 4.07697 | 193.225 | 0.12212 | 4.09433 | 193.307 |
| $Be_8H_{16}$ | 0.23437 | 2.13193 | 195.541 | 0.12104 | 4.13104 | 224.613 | 0.12066 | 4.14405 | 223.184 | 0.12115 | 4.12729 | 226.44 | 0.12073 | 4.14147 | 225.831 |
| $Be_9H_{18}$ | 0.2345 | 2.1312 | 223.019 | 0.12245 | 4.0833 | 256.736 | 0.12198 | 4.0992 | 255.455 | 0.12261 | 4.07814 | 259.486 | 0.12222 | 4.09115 | 259.188 |
| $Be_{10}H_{20}$ | 0.23391 | 2.1332 | 255.937 | 0.12261 | 4.07797 | 294.858 | 0.12134 | 4.12065 | 287.997 | 0.12184 | 4.10374 | 292.921 | 0.12141 | 4.11828 | 291.692 |

.[b] The hardness (η), softness (S) and polarizability (α) are in atomic units.

**Table 3.** Frontier molecular orbital energies and their gaps[c] of molecules computed at the various levels of theory.

| compd | HF | | | B3LYP | | | | | | B3P86 | | | | | |
|---|---|---|---|---|---|---|---|---|---|---|---|---|---|---|---|
| | 6-311++G** | | | 6-31+G** | | | 6-311++G** | | | 6-31+G** | | | 6-311++G** | | |
| | HOMO | LUMO | GAP | HOMO | LUMO | GAP | HOMO | LUMO | GAP | HOMO | LUMO | GAP | HOMO | LUMO | GAP |
| $BeH_2$ | -0.44838 | 0.03332 | 0.4817 | -0.31544 | -0.03782 | 0.27762 | -0.31871 | -0.0399 | 0.27881 | -0.33429 | -0.05108 | 0.28321 | -0.3371 | -0.05308 | 0.28402 |
| $Be_2H_4$ | -0.4456 | 0.00808 | 0.45368 | -0.3126 | -0.09307 | 0.21953 | -0.31607 | -0.09679 | 0.21928 | -0.3321 | -0.1097 | 0.2224 | -0.33501 | -0.11327 | 0.22174 |
| $Be_3H_6$ | -0.44758 | 0.02709 | 0.47467 | -0.31462 | -0.06615 | 0.24847 | -0.31694 | -0.06898 | 0.24796 | -0.33429 | -0.08383 | 0.25046 | -0.33636 | -0.08653 | 0.24983 |
| $Be_4H_8$ | -0.44703 | 0.2459 | 0.46293 | -0.31354 | -0.07901 | 0.23453 | -0.31575 | -0.08176 | 0.23399 | -0.33338 | -0.09787 | 0.23551 | -0.33535 | -0.10047 | 0.23488 |
| $Be_5H_{10}$ | -0.44633 | 0.0252 | 0.47153 | -0.31298 | -0.06789 | 0.24509 | -0.31486 | -0.07056 | 0.2443 | -0.33283 | -0.08678 | 0.24605 | -0.33452 | -0.08929 | 0.24523 |
| $Be_6H_{12}$ | -0.44569 | 0.02475 | 0.46844 | -0.31216 | -0.07256 | 0.2396 | -0.31409 | -0.07524 | 0.23885 | -0.33208 | -0.0921 | 0.23998 | -0.33381 | -0.09461 | 0.2392 |
| $Be_7H_{14}$ | -0.4452 | 0.02453 | 0.47073 | -0.3117 | -0.06704 | 0.24466 | -0.31358 | -0.06969 | 0.24389 | -0.33163 | -0.08635 | 0.24528 | -0.3331 | -0.08886 | 0.24424 |
| $Be_8H_{16}$ | -0.44484 | 0.2439 | 0.46874 | -0.31125 | -0.06918 | 0.24207 | -0.31315 | -0.07184 | 0.24131 | -0.33124 | -0.08895 | 0.24229 | -0.3329 | -0.09144 | 0.24146 |
| $Be_9H_{18}$ | -0.44456 | 0.02444 | 0.469 | -0.31112 | -0.06622 | 0.2449 | -0.31284 | -0.06889 | 0.24395 | -0.3309 | -0.08569 | 0.24521 | -0.33262 | -0.08819 | 0.24443 |
| $Be_{10}H_{20}$ | -0.40221 | 0.01562 | 0.4683 | -0.31642 | -0.0712 | 0.24522 | -0.31264 | -0.06996 | 0.24268 | -0.33077 | -0.08709 | 0.24368 | -0.33241 | -0.08959 | 0.24282 |

.[c] The frontier molecular orbital energies and their gaps are in atomic units.

Table 4. The Mulliken atomic charges of molecules computed from B3LYP/6-311++G** level.

| Monomer | Dimer | Trimer | Tetramer | Pentamer | Hexamer | Heptamer | Octamer | Nonamer | Decamer |
|---|---|---|---|---|---|---|---|---|---|
| H -0.148 | H -0.182 | H -0.137 | H -0.131 | H -0.143 | H -0.128 | H -0.135 | H -0.146 | H -0.134 | H -0.141 |
| Be 0.296 | Be 0.300 | Be 0.219 | Be 0.812 | Be 0.801 | Be 0.823 | Be 0.880 | Be 0.975 | Be 0.885 | Be 0.943 |
| H -0.148 | H -0.118 | H -0.127 | H -0.117 | H -0.116 | H -0.116 | H -0.117 | H -0.117 | H -0.117 | H -0.117 |
| | H -0.118 | H -0.126 | H -0.117 | H -0.116 | H -0.116 | H -0.116 | H -0.117 | H -0.116 | H -0.117 |
| | Be 0.300 | Be 0.343 | Be -0.310 | Be 0.411 | Be -0.062 | Be 0.239 | Be 0.747 | Be 0.106 | Be 0.423 |
| | H -0.182 | H -0.126 | H -0.135 | H -0.127 | H -0.125 | H -0.127 | H -0.127 | H -0.127 | H -0.127 |
| | | H -0.127 | H -0.134 | H -0.127 | H -0.126 | H -0.126 | H -0.127 | H -0.127 | H -0.128 |
| | | Be 0.219 | Be -0.309 | Be -1.159 | Be -0.029 | Be -0.477 | Be -0.947 | Be -0.356 | Be -0.752 |
| | | H -0.137 | H -0.117 | H -0.129 | H -0.117 | H -0.115 | H -0.116 | H -0.117 | H -0.117 |
| | | | H -0.117 | H -0.129 | H -0.117 | H -0.115 | H -0.116 | H -0.117 | H -0.117 |
| | | | Be 0.812 | Be 0.411 | Be -0.028 | Be 0.424 | Be 0.210 | Be 0.425 | Be 0.398 |
| | | | H -0.131 | H -0.116 | H -0.125 | H -0.115 | H -0.114 | H -0.114 | H -0.115 |
| | | | | H -0.116 | H -0.126 | H -0.115 | H -0.114 | H -0.115 | H -0.115 |
| | | | | Be 0.801 | Be -0.063 | Be -0.477 | Be 0.209 | Be 0.056 | Be 0.201 |
| | | | | H -0.143 | H -0.116 | H -0.127 | H -0.116 | H -0.115 | H -0.116 |
| | | | | | H -0.116 | H -0.126 | H -0.116 | H -0.115 | H -0.116 |
| | | | | | Be 0.822 | Be 0.239 | Be -0.947 | Be 0.426 | Be 0.201 |
| | | | | | H -0.128 | H -0.117 | H -0.127 | H -0.117 | H -0.115 |
| | | | | | | H -0.117 | H -0.127 | H -0.117 | H -0.115 |
| | | | | | | Be 0.880 | Be 0.975 | Be -0.364 | Be 0.398 |
| | | | | | | H -0.135 | H -0.117 | H -0.129 | H -0.117 |
| | | | | | | | H -0.117 | H -0.125 | H -0.117 |
| | | | | | | | Be 0.747 | Be 0.114 | Be -0.753 |
| | | | | | | | H -0.146 | H -0.117 | H -0.127 |
| | | | | | | | | H -0.116 | H -0.128 |
| | | | | | | | | Be 0.884 | Be 0.424 |
| | | | | | | | | H -0.134 | H -0.117 |
| | | | | | | | | | H -0.117 |
| | | | | | | | | | Be 0.944 |
| | | | | | | | | | H -0.141 |

5

**Table 5.** Calculated Mulliken charges, bond populations and bond lengths of crystalline $BeH_2$

| Species | S | P | Total | Charge(e) | Bonds | Population | Length(A) |
|---|---|---|---|---|---|---|---|
| **H(1)** | 1.23 | 0 | 1.23 | -0.23 | Be(1)-H(1) | 0.58 | 1.435 |
| **H(2)** | 1.24 | 0 | 1.24 | -0.24 | Be(2)-H(1) | 0.55 | 1.437 |
| **Be(1)** | 0.27 | 1.23 | 1.50 | 0.50 | Be(2)-H(2) | 0.57 | 1.44 |
| **Be(2)** | 0.28 | 1.27 | 1.55 | 0.45 | | | |

**Table 6.** The IR spectral characteristic of oligomers computed from the B3LYP/6-311++G** level.

| Compd | Frequencies, $cm^{-1}$ (infrared intensities, km/mol) |
|---|---|
| $BeH_2$ | **740(640.2156)**, 2260(155.30859) |
| $Be_2H_4$ | 330(233.5315), 590(96.13397), 950(259.55638), **1480(583.10452)** ,1590(169.05777), 2150(401.73924) |
| $Be_3H_6$ | 480(121.97897), 580(146.84479), 880(176.91711), **1510(924.18356)**, 1620(242.5545), 2150(337.63942) |
| $Be_4H_8$ | 480(223.2393), 590(96.12613), 890(195.85849), **1470(1007.0334), 1520(1355.53286)**, 1610(321.1884), 2150(470.11557) |
| $Be_5H_{10}$ | 500(162.62807), 590(129.5057), 890(130.90608), **1480(3423.37383), 1530(328.28617)**, 1610(643.69), 2150(449.44805) |
| $Be_6H_{12}$ | 490(136.37086), 590(104.18783), 890(176.59815), **1460(3818.97489), 1480(2203.34597)**, 1610(760.4181), 2150(487.83535) |
| $Be_7H_{14}$ | 500(155.97063), 590(107.64053), 890(137.27759), **1450(5534.99766)**, 1500(280.30368), 1610(785.0372), 2150(484.4817) |
| $Be_8H_{16}$ | 500(218.07), 590(139.46), 840(67.354), 890(141.35), **1440(7298.2)**, 1560(72.374), **1610(1014.1)**, 2150(493.83) |
| $Be_9H_{18}$ | 500(152.01027), 590(113.50592), 840(85.21636), 890(141.66471), **1430(6916.35851)**, 1540(56.27978), 1560(86.78147), **1610(1054.39831)**, 1770(30.23255), 2150(488.87033) |
| $Be_{10}H_{20}$ | 500(207.87278), 590(127.44316), 840(105.87406), 890(135.34063), **1430(7089.63023)**, 1540(114.39004), **1610(1218.78971)**, 2150(494.21322) |

Table 7. Raman spectral characteristic of oligomers computed from the B3LYP/6-311++G** level.

| Compd | Frequencies, $cm^{-1}$ (Raman activity) |
|---|---|
| $BeH_2$ | 2060(17.24), 2070(83.07), 2080(99.77), 2090(19.25) |
| $Be_2H_4$ | 720(20.40), 1470(12.19), 1790(10.29), 1800(59.99), 1810(20.62), 2180(18.05), 2190(78.32), 2200(153.97), 2210(27.93) |
| $Be_3H_6$ | 510(14.72), 690(15.51), 1460(10.48), 1790(15.20), 1800(74.41), 1810(68.79), 1820(13.52), 2170(18.23), 2180(70.20), 2190(233.79), 2200(40.32), 2210(13.16) |
| $Be_4H_8$ | 390(18.23), 780(18.64), 790(13.19), 1460(15.65), 1470(13.52), 1790(28.61), 1800(71.65), 1810(135.06), 1820(24.09), 2160(12.17), 2170(33.87), 2180(188.28), 2190(105.01), 2200(23.72) |
| $Be_5H_{10}$ | 310(11.67), 810(29.92), 820(15.00), 1460(11.86), 1470(28.54), 1780(11.36), 1790(38.11), 1800(88.53), 1810(195.62), 1820(35.06), 1830(11.84), 2160(16.29), 2170(53.05), 2180(289.45), 2190(65.86), 2200(18.50) |
| $Be_6H_{12}$ | 260(12.14), 820(15.89), 830(40.32), 1460(14.87), 1470(34.90), 1600(10.09), 1780(14.89), 1790(53.30), 1800(110.36), 1810(243.02), 1820(44.46), 1830(14.83), 2160(19.50), 2170(70.56), 2180(294.37), 2190(52.80), 2200(16.45) |
| $Be_7H_{14}$ | 830(26.86), 840(33.80), 1460(16.21), 1470(45.69), 1780(18.07), 1790(65.96), 1800(129.53), 1810(293.17), 1820(54.24), 1830(17.94), 2160(21.52), 2170(82.57), 2180(277.28), 2190(47.76), 2200(15.60) |
| $Be_8H_{16}$ | 200(11.54), 830(12.50), 840(68.15), 850(16.42), 1460(18.47), 1470(55.11), 1480(11.48), 1780(21.34), 1790(79.31), 1800(146.00), 1810(344.29), 1820(64.19), 1830(21.10), 1840(10.30), 2160(22.47), 2170(88.08), 2180(270.97), 2190(46.49), 2200(15.43) |
| $Be_9H_{18}$ | 840(45.48), 850(33.87), 1460(20.74), 1470(65.82), 1480(13.60), 1770(11.49), 1780(24.74), 1790(93.49), 1800(163.20), 1810(395.45), 1820(73.129), 1830(24.09), 1840(11.77), 2150(10.18), 2160(24.08), 2170(98.83), 2180(250.52), 2190(43.31), 2200(14.82) |
| $Be_{10}H_{20}$ | 740(12.31), 840(46.22), 850(31.08), 1130(11.65), 1460(21.69), 1470(53.22), 1480(10.25), 1590(11.53), 1600(14.94), 1610(29.26), 1630(13.05), 1770(12.82), 1780(30.19), 1790(124.46), 1800(190.40), 1810(264.46), 1820(50.63), 1830(20.36), 1840(10.45), 1890(14.23), 1900(47.90), 1910(182.65), 1920(32.83), 1930(10.86), 2150(14.15), 2160(40.26), 2170(211.81), 2180(159.78), 2190(31.99), 2200(12.26) |